\documentclass{revtex4-1}

\usepackage{amsmath}
\usepackage{amsthm}
\usepackage{graphicx}

\newcommand\ket[1]{\left|#1\right>}
\newcommand\bra[1]{\left<#1\right|}

\newcommand\expect[1]{\left<#1\right>}

\newcommand{\Mean}{\mathbf{M}}
\newcommand\half{\frac{1}{2}}

\newcommand{\Schr}{Schr\"odinger}

\newcommand\e{\mathrm{e}}

\newcommand\ro{{\hat\rho}}
\newcommand\Ho{{\hat H}}
\newcommand\Vo{{\hat V}}

\newcommand\denso{{\hat\varrho}}

\newcommand\rv{\mathbf{r}}
\newcommand\sv{\mathbf{s}}
\newcommand\xv{\mathbf{x}}
\newcommand{\pv}{\mathbf{p}}
\newcommand{\xvo}{\hat\xv}
\newcommand{\pvo}{\hat\pv}
\newcommand{\xve}{\expect{\xvo}}
\newcommand{\pve}{\expect{\pvo}}
\newcommand{\xvoc}{\xvo_c}
\newcommand{\wv}{\mathbf{w}}
\newcommand{\Dx}{\Delta\mathrm{x}}
\newcommand{\omG}{\omega_G}
\newcommand{\VPsi}{\Vo_\Psi}

\newcommand\Dcal{\mathcal{D}}

\begin{document}
\title{\Schr--Newton equation with spontaneous wave function collapse}
\author{Lajos Di\'osi}
\email{diosi.lajos@wigner.hu}
\affiliation{\quad Wigner Research Centre for Physics, 
                      H-1525 Budapest 114, P. O. Box 49, Hungary\\
                      and\\
                      \quad E\"otv\"os Lor\'and University, H-1117 Budapest, 
                      P\'azm\'any P\'eter stny. 1/A}
\date{\today}

\begin{abstract}
Based on the assumption that the standard \Schr~equation  becomes 
gravitationally modified for massive macroscopic objects, two independent proposals
has survived from the nineteen-eighties. The \Schr--Newton equation (1984)
provides well-localized solitons for free macro-objects but 
lacks the mechanism how extended wave functions collapse on solitons.
The gravity-related stochastic \Schr~equation (1989) provides the spontaneous
collapse  but the resulting solitons undergo a tiny diffusion leading to
an inconvenient steady increase of the kinetic energy.
We propose the stochastic \Schr--Newton equation which contains
the above two gravity-related modifications together.  
Then the wave functions of free macroscopic bodies will gradually and 
stochastically collapse to solitons which perform inertial motion
without the momentum diffusion: conservation of momentum  and
energy is restored. 
\end{abstract}

\maketitle

\section{Introduction}
The conjectured validity of quantum theory in  the macroscopic world
is famously problematic. This  is shown usually on the 
elementary example: free motion of an isolated  macroscopic mass $M$.
Its center of mass (c.o.m) wave packet is widening eternally
while we would expect a certain stationary localization.
The most spectacular paradox concerns the \Schr~cat states which are
superpositions of two distant wave packets, legitimate in
the microworld but problematic for a macroscopic body.  

A Newtonian semiclassical (mean-field) modification of the \Schr~equation to ensure 
stationary localization  was proposed first \cite{diosi1984}. 
This \Schr--Newton equation (SNE) yields well-localized soliton
wave packets for the mass $M$.  However, the   \Schr~cat  states remain 
legitimate solutions and an independent mechanism is required to destroy them --- 
as stated in ref. \cite{diosi1984}.
To this end, another Newton-gravity-related (G-related) mechanism was proposed
in two steps. 
The  G-related master equation (G-ME) decoheres the cat states \cite{diosi1987}.
Its unraveling,
the G-related stochastic  \Schr~equation (G-SSE) \cite{diosi1989} collapses them 
spontaneously  to one 
or the other wave packet, and drives this component into a soliton --- with one 
annoying effect though. The c.o.m. motion of the soliton 
never becomes stationary, it remains subject of tiny stochastic fluctuations
generating kinetic energy at constant rate.

Here we propose the combination of the SNE and the 
G-SSE, obtaining the new stochastic SNE (SSNE). 
We show that the momentum diffusion of solitons of 
massive isolated objects get canceled by the interplay between the stochasticity
of the collapse mechanism and the semiclassical Newtonian self-interaction. 
The remainder of the stochasticity  is merely an extreme small universal 
coordinate diffusion proportional to $\hbar/M$. 
Our SSNE is partially realizing the concept of  ref.~\cite{diosi2019}
where the diffusion effects of a stereotypical collapse equation 
[our eq. (\ref{SSEdx}) below]  
were completely eliminated by  imposing frame drag.  Also a related vision of 
induced gravity was put forward. 

Secs.~\ref{SecSNE} and \ref{SecSSE} recapitulate the basics of the SNE and the
G-SSE, invoking the results  of  
Refs.~\cite{diosi1984,diosi1987,diosi1989}. 
The combination of the two equations is introduced in sec.~\ref{SecSSNE}.
Then sec.~\ref{SecSoliton}
proves that the soliton momenta of SSNE   
are constant, their diffusion is canceled.   
Sec.~\ref{Final} contains final remarks.

\section{Semiclassical \Schr--Newton Equation}\label{SecSNE}
Consider the standard \Schr~equation 
$d\ket{\Psi}/dt=-(i/\hbar)\Ho\ket{\Psi}$
of a massive non-relativistic many-body system, with Hamiltonian $\Ho$
not including the Newtonian pair potential. Let $\denso(\rv)$ denote the 
operator of mass density at location $\rv$,
normalized to the total mass $M$. Instead of the Newton pair potential,
one postulates that the classical gravitational  is sourced by the 
mean-field $\expect{\denso(\sv)}=\bra{\Psi}\denso(\rv)\ket{\Psi}$.
The semiclassical SNE reads
\begin{eqnarray}\label{SNE}
\frac{d\ket{\Psi}}{dt}&=&-\frac{i}{\hbar}\Ho\ket{\Psi}+\frac{i}{\hbar}G
\int\int\frac{\denso(\rv)\expect{\denso(\sv)}d\rv d\sv}{\vert\rv-\sv\vert}\ket{\Psi}
\nonumber\\
                                        &\equiv&-\frac{i}{\hbar}(\Ho+\VPsi)\ket{\Psi},  
\end{eqnarray}
where $\VPsi$ is the $\Psi$ dependent Newtonian 
semiclassical (mean-field)  interaction.

The  SNE possesses soliton solutions for the c.o.m. wave function of isolated bodies. 
Consider the motion of a single spherical symmetric rigid mass $M$,  
the canonical position and momentum operators are $\xvo,\pvo$ respectively. 
The mass density operator can be written as
\begin{equation}\label{mdens}
\denso(\rv)=M f(\rv-\xvo)
\end{equation}
where $f$ is a normalized non-negative spherical symmetric function. 
Let us introduce the frequency parameter $\omG$  
as defined in ref.~\cite{diosi2007}:
\begin{equation}\label{omG}
\omG^2=\frac{4\pi}{3}GM\int f^2(\rv)d\rv.
\end{equation}
This sets an effective strength of Newtonian self-attraction
(of the G-related spontaneous collapse, too, cf.  
secs. \ref{SecSSE}-\ref{SecSoliton}) 
when  the position uncertainty $\Dx$ in state $\ket{\Psi}$ is
much smaller than the characteristic length scale(s) of $f(\rv)$ \cite{note}.
Then, ignoring higher (than 2nd) order terms in $\xvoc=\xvo-\xve$, we can write 
the SNE (\ref{SNE}) into   the simple form (cf. Appendix \ref{A}):
\begin{equation}\label{SNEdx}
\frac{d\ket{\Psi}}{dt}=-{\frac{i}{\hbar}}\Ho\ket{\Psi}
-\frac{i}{2\hbar}M\omG^2\xvoc^2\ket{\Psi}.
\end{equation}
If $\Ho=\pvo^2/2M$ and $\xve=0$ 
then the ground state coincides with the ground state of a central 
harmonic oscillator of frequency $\omG$:
\begin{equation}\label{SNEsol}
\Psi_0(\xv)=\mathcal{N}\exp
\left(-\frac{M\omG\xv^2}{2\hbar}\right).
\end{equation}
This is a soliton standing at the origin.
The soliton and its inertially traveling versions are  suitably
representing the stationary localization of free macro objects. 

However, there is no mechanism that drives larger solitons and large macroscopically
extended wave functions toward  the basic soliton states of shape (\ref{SNEsol}).
Consider, e.g., a \Schr~cat state which is superposition of two solitons 
at a large distance $\ell$ from each other:
\begin{equation}\label{cat}
\ket{\mbox{Cat}}=\frac{\ket{\mbox{left}}+\ket{\mbox{right}}}{\sqrt{2}}.
\end{equation}
They attract each other via $\VPsi$,
can even 
form a Kepler system, but cannot collapse to a single soliton. 
Ref. \cite{diosi1984} emphasizes that collapse needs an additional
mechanism, not present in the SNE.

\section{Gravity-Related Wavefunction Collapse}\label{SecSSE}
Alternatively to the SNE (\ref{SNE}), 
the G-SSE
considers the following stochastic modification of the standard \Schr~equation: 
\begin{equation}\label{SSE}
\frac{d\ket{\Psi}}{dt}=-\frac{i}{\hbar}\Ho\ket{\Psi}
-\frac{G}{2\hbar}\int\int\frac{\denso_c(\rv)\denso_c(\sv)d\rv d\sv}{\vert\rv-\sv\vert}\ket{\Psi}
+\frac{1}{\hbar}\int\denso_c(\rv)\Phi(\rv)d\rv\ket{\Psi},
\end{equation}
where $\denso_c(\rv)=\denso(\rv)-\expect{\denso(\rv)}$ 
and $\Phi(\rv,t)$ is a white-noise field of spatial correlation 
\begin{equation}\label{PhiPhi}
\Mean\Phi(\rv,t)\Phi(\sv,\tau)=\frac{G\hbar}{\vert\rv-\sv\vert}\delta(t-\tau),
\end{equation}
$\Mean$ stands for the stochastic mean.
The white-noise is to be taken in terms of the Ito differential calculus.
This G-SSE
yields the spontaneous collapse of massive macroscopic spatial superpositions,
of the \Schr~cat states in particular. 
The average state, i.e., the density matrix $\ro=\Mean\ket{\Psi}\bra{\Psi}$
satisfies the G-ME:
\begin{eqnarray}\label{ME}
\frac{d\ro}{dt}&=&-\frac{i}{\hbar}[\Ho,\ro]
-\frac{G}{2\hbar}
\int\int\frac{[\denso(\rv),[\denso(\sv),\ro]]d\rv d\sv}{\vert\rv-\sv\vert}\nonumber\\
&\equiv&-\frac{i}{\hbar}[\Ho,\ro]+\Dcal\ro.
\end{eqnarray}
This suppresses the coherence of macroscopically
distinct superpositions of the mass density, yielding their statistical 
mixture without the collapse.
The G-SSE (\ref{SSE}) adds the collapse as well to the decoherence.  
The two distant wavepackets of the  \Schr~cat state (\ref{cat})
collapse onto one of them randomly at the rate
\begin{equation}\label{DP}
\frac{\Delta E_G}{\hbar},
\end{equation}
where $\Delta E_G$ denotes how much the collapse reduces the gravitational 
energy of the cat \cite{penrose1996,penrose1998,penrose2014}.

The G-SSE (\ref{SSE}) becomes simple for a single
mass in the regime of small $\Dx$ (cf. Appendices \ref{B},\ref{C}): 
\begin{equation}\label{SSEdx}
\frac{d\ket{\Psi}}{dt}=-\frac{i}{\hbar}\Ho\ket{\Psi}
-\frac{1}{2\hbar}M\omG^2\xvoc^2\ket{\Psi}
+\sqrt{\frac{M}{\hbar}}\omG\xvoc\wv\ket{\Psi},
\end{equation}
where $\wv$ is the vector of three independent standard white-noises.
(Correspondence with notations in ref.~\cite{diosi1989} is 
$\Gamma=2M\omG^2/\hbar$  and $d\xi=\sqrt{M/\hbar}\omG\wv$.)
Note, the Hermitian self-attracting potential in the SNE (\ref{SNEdx})
becomes anti-Hermitian here.
For free bodies ($\Ho=\pvo^2/2M$), the solutions converge to
solitons of the steady shape
\begin{equation}\label{SSEsol}
\Psi_0(\xv)=\mathcal{N}\exp
\left(-(1-i)\frac{M\omG\xv^2}{2\hbar}\right).
\end{equation}
This is slightly different from (\ref{SNEsol}), by the new complex factor 
$1-i$ which leaves the spread $\Dx=\sqrt{(\hbar/2M\omG})$ unchanged 
just creating the  correlation $\hbar/2$ between $\xvo$ and $\pvo$. 
But the major difference is in the c.o.m. motion which is plagued by a 
certain correlated diffusion of both the position and the momentum:
\begin{eqnarray}
\frac{d}{dt}\xve&=&\frac{\pve}{M}+\sqrt{\hbar/M}\wv,\label{xdiff}\label{dotx}\\
\frac{d}{dt}\pve&=&\sqrt{\hbar M}\omG\wv.\label{pdiff}\label{dotp}
\end{eqnarray}

The  momentum diffusion (\ref{pdiff}) is increasing the kinetic energy at rate $\half\hbar\omG^2$    
which is probably an unphysical artifact of the collapse model.
We get rid of it below..
 
\section{\Schr--Newton Equation with Wavefunction Collapse}\label{SecSSNE}
We slightly alter the original 1989-version (\ref{SSE}) of the G-SSE.
We insert a factor $\exp(-i\pi/4)=(1-i)/\sqrt{2}$ in front of the anti-Hermitian
stochastic potential:  
\begin{equation}\label{SSEi}
\frac{d\ket{\Psi}}{dt}=-\frac{i}{\hbar}\Ho\ket{\Psi}
-\frac{G}{2\hbar}\int\int\frac{\denso_c(\rv)\denso_c(\sv)d\rv d\sv}{\vert\rv-\sv\vert}\ket{\Psi}
+\frac{\e^{-i\pi/4}}{\hbar}\int\denso_c(\rv)\Phi(\rv)d\rv\ket{\Psi}.
\end{equation}
This yields the same G-ME (\ref{ME}) for the density matrix, and also encodes
the collapse of macroscopic superpositions. The novel feature is the appearance
of the Hermitian stochastic potential $\Phi(\rv)/\sqrt{2}$ which will cancel
the momentum diffusion (\ref{pdiff}) of the solitons ---
provided we include the gravitational self-attraction contained 
in $\Vo_\Psi$ of the SNE.

Our new proposal is the following combination of the 
SNE (\ref{SNE}) and the modified G-SSE (\ref{SSEi}):
\begin{equation}\label{SSNE}
\frac{d\ket{\Psi}}{dt}=-\frac{i}{\hbar}(\Ho\!+\!\VPsi)\!\ket{\Psi}
-\frac{G}{2\hbar}\int\int\frac{\denso_c(\rv)\denso_c(\sv)d\rv d\sv}{\vert\rv-\sv\vert}\ket{\Psi}
+\frac{\e^{-i\pi/4}}{\hbar}\!\!\int\!\!\denso_c(\rv)\Phi(\rv)d\rv\!\ket{\Psi}\!.
\end{equation}
The ME of the average state reads:
\begin{equation}\label{ME22}
\frac{d\ro}{dt}=-\frac{i}{\hbar}[\Ho+\VPsi,\ro]+\Dcal\ro.
\end{equation}
This is not a closed equation for $\ro$ since $\VPsi$ depends on the
pure state $\ket{\Psi}$. The lack of a closed linear ME is the signature
of anomalies \cite{gisin1989}  already troubling  the SNE  
and inherited by our SSNE (\ref{SSNE}). There is, however, a difference.
The spontaneous collapse might shadow the anomalies.  
One of them, the fake action-at-a-distance is based on the attraction 
caused by $\VPsi$ between the  two halves 
$\ket{\mbox{left}},\ket{\mbox{right}}$ of the \Schr~cat \cite{diosi2016}.
Unlike in case of the SNE, the cat now has the finite lifetime $\hbar/\Delta E_G$
which may be 
too short to reach detectable shifts of the left or the right wavepackets
\cite{grossardt2022}. 

\section{Solitons with Energy Conservation}\label{SecSoliton}
We consider the soliton solutions of the new  
spontaneous collapse dynamics SSNE
in the small-$\Dx$ approximation. 
The single body special case of the SSNE (\ref{SSNE})
takes this form:  
\begin{equation}\label{SSNEdx}
\frac{d\ket{\Psi}}{dt}=-{\frac{i}{\hbar}}\Ho\ket{\Psi}
-\frac{1+i}{2\hbar}M\omG^2\xvoc^2\ket{\Psi}
+(1-i)\sqrt{\frac{M}{2\hbar}}\omG\xvoc\wv\ket{\Psi}.
\end{equation}
Note the important complex factors $(1+i)$ and $(1-i)$ compared
to the small-$\Dx$ approximation (\ref{SSEdx}) of the G-SSE model.  
With these two factors, as we show below, 
the soliton's momentum diffusion (\ref{dotp}) cancels. 

For the time-dependent wave function of the soliton we
take the following Ansatz:
\begin{eqnarray}\label{Ansatz}
\Psi_t(\xv)&=&\mathcal{N}\exp
\left(-(1-i)\frac{\vert\xv-\xve_t\vert^2}{4\Dx^2}
          +\frac{i}{\hbar}\pve\xvo\right),\nonumber\\
\Dx^2&=&\frac{\hbar}{\sqrt{2}M\omG},\nonumber\\
\frac{d}{dt}\xve_t&=&\frac{\pv}{M}+\sqrt{\hbar/M}\wv_t.
\end{eqnarray}
These solutions correspond to inertial c.o.m. motion at constant momentum $\pve$,
apart from a minuscule diffusion of the coordinate.     

We still owe to show that the wavefunciton $\Psi_t(\xv)$
satisfies the eq. (\ref{SSNEdx}). It is sufficient if we prove it for the
soliton initially at rest at the origin, i.e., for $\xve_0=0$ and $\pve=0$.
First, we apply the eq. (\ref{SSNEdx}) to $\Psi_0(\xv)$:
\begin{equation}\label{dPsiSSE}
\frac{d\Psi_0(\xv)}{dt}=
\left(\!i\hbar\frac{\nabla^2}{2M}
\!-\!\frac{1+i}{2\hbar}M\omG^2\xv^2
\!+\!(1\!-\!i)\sqrt{\frac{M}{2\hbar}}\omG\xv\wv\!\right)\!\!\Psi_0(\xv).\nonumber
\end{equation}
Second, we derive $d\Psi_0(\xv)/dt$ from the Ansatz (\ref{Ansatz}):
\begin{equation}\label{dPsi}
\frac{d\Psi_0(\xv)}{dt}=\left(
-\sqrt{\hbar/M}\wv\nabla + \frac{\hbar}{2M}\nabla^2\right)\Psi_0(\xv), 
\end{equation}
where the second term is comes from the Ito-correction
$(\xv\wv dt\nabla)^2=\xv^2\nabla^2 dt$.  
We substitute the expression
$$\nabla\Psi_0=-(1-i)\frac{\xv}{2\Dx^2}\Psi_0
                             =-(1-i)\frac{\xv}{2}\frac{\sqrt{2}M\omG}{\hbar}\Psi_0
$$ 
and then observe that the stochastic term coincides with that of (\ref{dPsiSSE}).
We also substitute the expression
$$
\nabla^2\Psi_0
=\left( -i\frac{\xv^2}{2\Dx^4} -(1-i)\frac{1}{2\Dx^2}\right)\Psi_0
$$
in both eq. (\ref{dPsiSSE}) and eq. (\ref{dPsi}). Then, after elementary algebraic
steps, their deterministic parts will also coincide.

\section{Final remarks}\label{Final}
Penrose also proposed the spontaneous 
collapse rate (\ref{DP}) of the \Schr~cat as well as the SNE (\ref{SNE}) to generate the 
stationary states after the collapse  \cite{penrose1996,penrose1998,penrose2014}.
He is treating the G-related spontaneous collapse and the SNE
together.  However, he has been in a holding position
regarding any concrete dynamics (like the G-SSE or others) 
of the collapse. 

With the goal of reaching a closed linear ME to
avoid the anomalies of the SNE, ref.~\cite{nimmrichter2015}
proposed a formal completion of the SNE by stochastic terms.
Ref.~\cite{tilloydiosi2016} showed that this goal can be achieved
by combining the SNE with the G-SSE, of course differently
from the present proposal SSNE which, contrary to 
refs.~\cite{nimmrichter2015, tilloydiosi2016},
sacrifices the closed ME in favor of
energy-momentum conservation --- at least in the free c.o.m.
motion of the macroscopic body.   
 
The advantage of the newly proposed SSNE (\ref{SSNE}) compared to the
G-SSE (\ref{SSE}) is  this: 
when the state of a  free massive body is collapsing towards the
localized soliton, the spontaneous gain of kinetic energy is 
gradually disappearing and
in the soliton states the energy-momentum conservation becomes restored.   
The non-conservations by the G-SSE (and by 
other collapse models) are probably warnings of infancy of the models. 
Yet, their related predictions are the only testable effects currently \cite{bassi2013},
since massive  \Schr~cat states are not available in the laboratory to date.
To test the G-SSE, ref.~\cite{helou2017} searched for the predicted c.o.m. 
momentum diffusion
in the super-precise data of Lisa Pathfinder experiment. 
Work \cite{donadi2021} was hunting the spontaneous radiation 
predicted by the momentum diffusion of the nuclei inside the detector
in the super-low-background Gran Sasso laboratory. Both
works put an upper bound on the strength of the G-related 
spontaneous collapse. If we adopt the new SSNE,
a different interpretation of the Lisa Pathfinder data is possible.
But the interpretation of the Gran Sasso data can be retained 
because the new model SSNE has not removed the momentum diffusion
of the microscopic constituents but of the c.o.m. of the macro-object.

Our proposal is the first dynamical model of spontaneous collapse with
partial restoration of energy-momentum conservation at the price
of typical anomalies of semiclassical theories which might become
partially masked by the collapse mechanism. Future works should
aim at more complete restoration of energy-momentum
conservation and at better understanding whether the said anomalies
would become completely neutralized.

This research was funded by 
the Foundational Questions Institute and Fetzer Franklin
Fund, a donor advised fund of Silicon Valley Community
Foundation (Grant No. FQXi-RFP-CPW-2008, and a mini-grant), 
the National Research, Development and Innovation Office
for ``Frontline'' Research Excellence Program (Grant No.
KKP133827), research grant (Grant. No. K12435),
and the John Templeton Foundation (Grant 62099).

\section*{A.}\label{A}

To derive $\Vo_\Psi$ of Eq. (\ref{SNE}) for small $\Dx$,
we start from
\begin{equation}\label{VPsi}
\VPsi=-GM^2
\int\int\frac{f(\rv-\xvo)\expect{f(\sv-\xvo)}d\rv d\sv}{\vert\rv-\sv\vert},
\end{equation}
and consider the expansions 
\begin{equation}
f(\rv-\xvo)=\left(1-(\xvoc\nabla)+\half(\xvoc\nabla)^2\right)f(\rv-\xve)
\expect{f(\rv-\xvo)}=\left(1+\half(\xvoc\nabla)^2\right)f(\rv-\xve)
\end{equation}
omitting higher order terms in $\xvoc$.
Translation invariance of $\VPsi$ allows us to set $\xve=0$.
\begin{equation}
f(\rv-\xvo)\expect{f(\sv-\xvo)}=
\left[1-\xvoc\nabla_r+\half(\xvoc\nabla_r)^2\right]f(\rv)
\left[1+\half(\xvoc\nabla_s)^2\right]f(\sv).
\end{equation}
Rotational invariance of $\VPsi$ cancels the linear term and yields
the identity $(\xvoc\nabla)^2=(1/3)\xvoc^2\Delta$, hence  
\begin{equation}
f(\rv-\xvo)\!\expect{f(\sv-\xvo)}\!=\!
f(\rv)f(\sv)\!+\!\frac{1}{6}\xvoc^2\left[f(\sv)\Delta f(\rv)\!+\!f(\rv)\Delta f(\sv)\right],
\end{equation}
ignoring higher orders of $\xvoc$.
Using this in Eq. (\ref{VPsi}):
\begin{eqnarray}
\VPsi&=&-GM^2
\int\int\frac{f(\rv)f(\sv)d\rv d\sv}{\vert\rv-\sv\vert}
+\frac{GM^2\xvoc^2}{6}
\int\int\frac{f(\sv)\Delta f(\rv)+f(\rv)\Delta f(\sv)}{\vert\rv-\sv\vert}d\rv d\sv
\nonumber\\
&=&2E_G+\frac{4\pi GM^2}{3}\xvoc^2\int f^2(\rv)d\rv=2E_G+\half M\omG^2\xvoc^2,
\end{eqnarray}
where partial integrations and the indetity 
$\Delta\vert \rv-\sv\vert^{-1}=-4\pi\delta(\rv-\sv)$
have been used. 
The constant $E_G$ stands for the gravitational self-energy.  
\section*{B}\label{B}
To derive the double integral in Eq. (\ref{SSE}) for small $\Dx$,
we start from
\begin{equation}
\frac{GM^2}{2}
\int\int\frac{f_c(\rv-\xvo)f_c(\sv-\xvoc) d\rv d\sv}{\vert\rv-\sv\vert},
\end{equation}
and substitute the expansion  
\begin{equation}
f_c(\rv-\xvo)=f(\rv-\xvo)-\expect{f(\rv-\xvo)}=-(\xvoc\nabla)f(\rv-\xve).
\end{equation}
Again, we can take $\xve=0$, yielding
\begin{equation}
\frac{GM^2\xvoc^2}{6}
\int\int\frac{\nabla_r f(\rv)\nabla_s f(\sv) d\rv d\sv}{\vert\rv-\sv\vert}
=\frac{4\pi GM^2\xvoc^2}{6}\int f^2(\rv)d\rv
=\half M\omG^2\xvoc^2.
\end{equation}
\section*{C}\label{C}
To derive the stochastic term in Eq. (\ref{SSE}) for small $\Dx$,
we write
\begin{eqnarray}
\frac{1}{\hbar}\int \denso_c(\rv)\Phi(\rv) d\rv=\frac{M}{\hbar}\xvo_c\int \nabla f(\rv)\Phi(\rv) d\rv
\end{eqnarray}
and introduce the new stochastic variable, linear in the old stochastic field $\Phi(\rv)$: 
\begin{equation}
\wv=\sqrt{\frac{M}{\hbar}}\omG^{-1}\int \nabla f(\rv-\xve)\Phi(\rv)d\rv.
\end{equation}
For the correlation function we obtain the following:
\begin{eqnarray}
\Mean \wv_t\circ\wv_\tau
&=&
\frac{M}{\hbar}\omG^{-2}\int\int(\nabla f(\rv)\circ\nabla f(\sv))\Mean\Phi(\rv,t)\Phi(\sv,\tau)d\rv d\sv\nonumber\\
&=&\frac{M}{\hbar}\omG^{-2}\int\int f(\rv)f(\sv)\nabla_r\circ\nabla_s)\frac{\hbar G}{\vert\rv-\sv\vert}d\rv d\sv\delta(t-\tau)\nonumber\\
&=&\omG^{-2}\frac{4\pi}{3}I_{3\times3} GM\int f^2(\rv)d\rv \delta(t-\tau)
=I_{3\times3}\delta(t-\tau)
\end{eqnarray}
where the $\Mean\Phi(\rv,t)\Phi(\sv,\tau)$ has been substituted by the expression (\ref{PhiPhi}).

\bibliography{SNE-DP}

\begin{thebibliography}{17}%
\makeatletter
\providecommand \@ifxundefined [1]{%
 \@ifx{#1\undefined}
}%
\providecommand \@ifnum [1]{%
 \ifnum #1\expandafter \@firstoftwo
 \else \expandafter \@secondoftwo
 \fi
}%
\providecommand \@ifx [1]{%
 \ifx #1\expandafter \@firstoftwo
 \else \expandafter \@secondoftwo
 \fi
}%
\providecommand \natexlab [1]{#1}%
\providecommand \enquote  [1]{``#1''}%
\providecommand \bibnamefont  [1]{#1}%
\providecommand \bibfnamefont [1]{#1}%
\providecommand \citenamefont [1]{#1}%
\providecommand \href@noop [0]{\@secondoftwo}%
\providecommand \href [0]{\begingroup \@sanitize@url \@href}%
\providecommand \@href[1]{\@@startlink{#1}\@@href}%
\providecommand \@@href[1]{\endgroup#1\@@endlink}%
\providecommand \@sanitize@url [0]{\catcode `\\12\catcode `\$12\catcode
  `\&12\catcode `\#12\catcode `\^12\catcode `\_12\catcode `\%12\relax}%
\providecommand \@@startlink[1]{}%
\providecommand \@@endlink[0]{}%
\providecommand \url  [0]{\begingroup\@sanitize@url \@url }%
\providecommand \@url [1]{\endgroup\@href {#1}{\urlprefix }}%
\providecommand \urlprefix  [0]{URL }%
\providecommand \Eprint [0]{\href }%
\providecommand \doibase [0]{http://dx.doi.org/}%
\providecommand \selectlanguage [0]{\@gobble}%
\providecommand \bibinfo  [0]{\@secondoftwo}%
\providecommand \bibfield  [0]{\@secondoftwo}%
\providecommand \translation [1]{[#1]}%
\providecommand \BibitemOpen [0]{}%
\providecommand \bibitemStop [0]{}%
\providecommand \bibitemNoStop [0]{.\EOS\space}%
\providecommand \EOS [0]{\spacefactor3000\relax}%
\providecommand \BibitemShut  [1]{\csname bibitem#1\endcsname}%
\let\auto@bib@innerbib\@empty
\bibitem [{\citenamefont {Di\'osi}(1984)}]{diosi1984}%
  \BibitemOpen
  \bibfield  {author} {\bibinfo {author} {\bibfnamefont {L.}~\bibnamefont
  {Di\'osi}},\ }\href {\doibase http://dx.doi.org/10.1016/0375-9601(84)90397-9}
  {\bibfield  {journal} {\bibinfo  {journal} {Physics Letters A}\ }\textbf
  {\bibinfo {volume} {105}},\ \bibinfo {pages} {199 } (\bibinfo {year}
  {1984})}\BibitemShut {NoStop}%
\bibitem [{\citenamefont {Di\'osi}(1987)}]{diosi1987}%
  \BibitemOpen
  \bibfield  {author} {\bibinfo {author} {\bibfnamefont {L.}~\bibnamefont
  {Di\'osi}},\ }\href@noop {} {\bibfield  {journal} {\bibinfo  {journal}
  {Physics letters A}\ }\textbf {\bibinfo {volume} {120}},\ \bibinfo {pages}
  {377} (\bibinfo {year} {1987})}\BibitemShut {NoStop}%
\bibitem [{\citenamefont {Di\'osi}(1989)}]{diosi1989}%
  \BibitemOpen
  \bibfield  {author} {\bibinfo {author} {\bibfnamefont {L.}~\bibnamefont
  {Di\'osi}},\ }\href@noop {} {\bibfield  {journal} {\bibinfo  {journal}
  {Physical Review A}\ }\textbf {\bibinfo {volume} {40}},\ \bibinfo {pages}
  {1165} (\bibinfo {year} {1989})}\BibitemShut {NoStop}%
\bibitem [{\citenamefont {Di{\'o}si}(2019)}]{diosi2019}%
  \BibitemOpen
  \bibfield  {author} {\bibinfo {author} {\bibfnamefont {L.}~\bibnamefont
  {Di{\'o}si}},\ }\href@noop {} {\bibfield  {journal} {\bibinfo  {journal}
  {Quantum Reports}\ }\textbf {\bibinfo {volume} {1}},\ \bibinfo {pages} {277}
  (\bibinfo {year} {2019})}\BibitemShut {NoStop}%
\bibitem [{\citenamefont {Di\'{o}si}(2007)}]{diosi2007}%
  \BibitemOpen
  \bibfield  {author} {\bibinfo {author} {\bibfnamefont {L.}~\bibnamefont
  {Di\'{o}si}},\ }\href {\doibase 10.1088/1751-8113/40/12/S07} {\bibfield
  {journal} {\bibinfo  {journal} {Journal of Physics A: Mathematical and
  Theoretical}\ }\textbf {\bibinfo {volume} {40}},\ \bibinfo {pages} {2989}
  (\bibinfo {year} {2007})}\BibitemShut {NoStop}%
\bibitem [{not()}]{note}%
  \BibitemOpen
  \href@noop {} {\ }\bibinfo {note} {The parameter $\omG$ is fully classical,
  has nothing to do with the quantum. It is in the $mHz$-range (weak G-related
  effects) if $f(\rv)$ does not resolve the microscopic structure. It can grow
  up to the $kHz$-range in case of deep subatomic resolution (strong G-related
  effects).}\BibitemShut {Stop}%
\bibitem [{\citenamefont {Penrose}(1996)}]{penrose1996}%
  \BibitemOpen
  \bibfield  {author} {\bibinfo {author} {\bibfnamefont {R.}~\bibnamefont
  {Penrose}},\ }\href@noop {} {\bibfield  {journal} {\bibinfo  {journal}
  {General relativity and gravitation}\ }\textbf {\bibinfo {volume} {28}},\
  \bibinfo {pages} {581} (\bibinfo {year} {1996})}\BibitemShut {NoStop}%
\bibitem [{\citenamefont {Penrose}(1998)}]{penrose1998}%
  \BibitemOpen
  \bibfield  {author} {\bibinfo {author} {\bibfnamefont {R.}~\bibnamefont
  {Penrose}},\ }\href@noop {} {\bibfield  {journal} {\bibinfo  {journal}
  {Philosophical transactions- Royal Society of London, Series A Mathematical
  Physical and Engineering Sciences}\ ,\ \bibinfo {pages} {1927}} (\bibinfo
  {year} {1998})}\BibitemShut {NoStop}%
\bibitem [{\citenamefont {Penrose}(2014)}]{penrose2014}%
  \BibitemOpen
  \bibfield  {author} {\bibinfo {author} {\bibfnamefont {R.}~\bibnamefont
  {Penrose}},\ }\href@noop {} {\bibfield  {journal} {\bibinfo  {journal}
  {Foundations of Physics}\ }\textbf {\bibinfo {volume} {44}},\ \bibinfo
  {pages} {557} (\bibinfo {year} {2014})}\BibitemShut {NoStop}%
\bibitem [{\citenamefont {Gisin}(1989)}]{gisin1989}%
  \BibitemOpen
  \bibfield  {author} {\bibinfo {author} {\bibfnamefont {N.}~\bibnamefont
  {Gisin}},\ }\href@noop {} {\bibfield  {journal} {\bibinfo  {journal} {Helv.
  Phys. Acta}\ }\textbf {\bibinfo {volume} {62}},\ \bibinfo {pages} {363}
  (\bibinfo {year} {1989})}\BibitemShut {NoStop}%
\bibitem [{\citenamefont {Di{\'o}si}(2016)}]{diosi2016}%
  \BibitemOpen
  \bibfield  {author} {\bibinfo {author} {\bibfnamefont {L.}~\bibnamefont
  {Di{\'o}si}},\ }in\ \href@noop {} {\emph {\bibinfo {booktitle} {Journal of
  Physics: Conference Series}}},\ Vol.\ \bibinfo {volume} {701}\ (\bibinfo
  {organization} {IOP Publishing},\ \bibinfo {year} {2016})\ p.\ \bibinfo
  {pages} {012019}\BibitemShut {NoStop}%
\bibitem [{\citenamefont {Gro{\ss}ardt}(2022)}]{grossardt2022}%
  \BibitemOpen
  \bibfield  {author} {\bibinfo {author} {\bibfnamefont {A.}~\bibnamefont
  {Gro{\ss}ardt}},\ }\href@noop {} {\bibfield  {journal} {\bibinfo  {journal}
  {AVS Quantum Science}\ }\textbf {\bibinfo {volume} {4}},\ \bibinfo {pages}
  {010502} (\bibinfo {year} {2022})}\BibitemShut {NoStop}%
\bibitem [{\citenamefont {Nimmrichter}\ and\ \citenamefont
  {Hornberger}(2015)}]{nimmrichter2015}%
  \BibitemOpen
  \bibfield  {author} {\bibinfo {author} {\bibfnamefont {S.}~\bibnamefont
  {Nimmrichter}}\ and\ \bibinfo {author} {\bibfnamefont {K.}~\bibnamefont
  {Hornberger}},\ }\href@noop {} {\bibfield  {journal} {\bibinfo  {journal}
  {Physical Review D}\ }\textbf {\bibinfo {volume} {91}},\ \bibinfo {pages}
  {024016} (\bibinfo {year} {2015})}\BibitemShut {NoStop}%
\bibitem [{\citenamefont {Tilloy}\ and\ \citenamefont
  {Di\'osi}(2016)}]{tilloydiosi2016}%
  \BibitemOpen
  \bibfield  {author} {\bibinfo {author} {\bibfnamefont {A.}~\bibnamefont
  {Tilloy}}\ and\ \bibinfo {author} {\bibfnamefont {L.}~\bibnamefont
  {Di\'osi}},\ }\href {\doibase 10.1103/PhysRevD.93.024026} {\bibfield
  {journal} {\bibinfo  {journal} {Phys. Rev. D}\ }\textbf {\bibinfo {volume}
  {93}},\ \bibinfo {pages} {024026} (\bibinfo {year} {2016})}\BibitemShut
  {NoStop}%
\bibitem [{\citenamefont {Bassi}\ \emph {et~al.}(2013)\citenamefont {Bassi},
  \citenamefont {Lochan}, \citenamefont {Satin}, \citenamefont {Singh},\ and\
  \citenamefont {Ulbricht}}]{bassi2013}%
  \BibitemOpen
  \bibfield  {author} {\bibinfo {author} {\bibfnamefont {A.}~\bibnamefont
  {Bassi}}, \bibinfo {author} {\bibfnamefont {K.}~\bibnamefont {Lochan}},
  \bibinfo {author} {\bibfnamefont {S.}~\bibnamefont {Satin}}, \bibinfo
  {author} {\bibfnamefont {T.}~\bibnamefont {Singh}}, \ and\ \bibinfo {author}
  {\bibfnamefont {H.}~\bibnamefont {Ulbricht}},\ }\href@noop {} {\bibfield
  {journal} {\bibinfo  {journal} {Reviews of Modern Physics}\ }\textbf
  {\bibinfo {volume} {85}},\ \bibinfo {pages} {471} (\bibinfo {year}
  {2013})}\BibitemShut {NoStop}%
\bibitem [{\citenamefont {Helou}\ \emph {et~al.}(2017)\citenamefont {Helou},
  \citenamefont {Slagmolen}, \citenamefont {McClelland},\ and\ \citenamefont
  {Chen}}]{helou2017}%
  \BibitemOpen
  \bibfield  {author} {\bibinfo {author} {\bibfnamefont {B.}~\bibnamefont
  {Helou}}, \bibinfo {author} {\bibfnamefont {B.}~\bibnamefont {Slagmolen}},
  \bibinfo {author} {\bibfnamefont {D.}~\bibnamefont {McClelland}}, \ and\
  \bibinfo {author} {\bibfnamefont {Y.}~\bibnamefont {Chen}},\ }\href@noop {}
  {\bibfield  {journal} {\bibinfo  {journal} {Physical Review D}\ }\textbf
  {\bibinfo {volume} {95}},\ \bibinfo {pages} {084054} (\bibinfo {year}
  {2017})}\BibitemShut {NoStop}%
\bibitem [{\citenamefont {Donadi}\ \emph {et~al.}(2021)\citenamefont {Donadi},
  \citenamefont {Piscicchia}, \citenamefont {Curceanu}, \citenamefont
  {Di{\'o}si}, \citenamefont {Laubenstein},\ and\ \citenamefont
  {Bassi}}]{donadi2021}%
  \BibitemOpen
  \bibfield  {author} {\bibinfo {author} {\bibfnamefont {S.}~\bibnamefont
  {Donadi}}, \bibinfo {author} {\bibfnamefont {K.}~\bibnamefont {Piscicchia}},
  \bibinfo {author} {\bibfnamefont {C.}~\bibnamefont {Curceanu}}, \bibinfo
  {author} {\bibfnamefont {L.}~\bibnamefont {Di{\'o}si}}, \bibinfo {author}
  {\bibfnamefont {M.}~\bibnamefont {Laubenstein}}, \ and\ \bibinfo {author}
  {\bibfnamefont {A.}~\bibnamefont {Bassi}},\ }\href@noop {} {\bibfield
  {journal} {\bibinfo  {journal} {Nature Physics}\ }\textbf {\bibinfo {volume}
  {17}},\ \bibinfo {pages} {74} (\bibinfo {year} {2021})}\BibitemShut {NoStop}%
\end{thebibliography}%
\end{document}